\documentstyle[11pt,newpasp,twoside,epsf]{article}
\def\edcomment#1{\iffalse\marginpar{\raggedright\sl#1\/}\else\relax\fi}
\reversemarginpar

\def\beq{\begin{equation}}
\def\eeq{\end{equation}}
\def\bey{\begin{eqnarray}}
\def\eey{\end{eqnarray}}

\def\mpc{\,h^{-1}{\rm {Mpc}}}
\def\kms{\,{\rm {km\, s^{-1}}}}

\def\zetarr{\zeta(r_{12},r_{23},r_{31})}

\def\wrp#1{w(r_{p#1})}
\def\gs{\mathrel{\raise1.16pt\hbox{$>$}\kern-7.0pt
\lower3.06pt\hbox{{$\scriptstyle \sim$}}}}
\def\ls{\mathrel{\raise1.16pt\hbox{$<$}\kern-7.0pt
\lower3.06pt\hbox{{$\scriptstyle \sim$}}}}
\def\gtsima{$\; \buildrel > \over \sim \;$}
\def\ltsima{$\; \buildrel < \over \sim \;$}
\def\prosima{$\; \buildrel \propto \over \sim \;$}
\def\gsim{\lower.5ex\hbox{\gtsima}}
\def\lsim{\lower.5ex\hbox{\ltsima}}
\def\simgt{\lower.5ex\hbox{\gtsima}}
\def\simlt{\lower.5ex\hbox{\ltsima}}
\def\simpr{\lower.5ex\hbox{\prosima}}

\begin{document}
\title {
Slow relative motion of IRAS galaxies at small separations: 
implications for galaxy formation models}
\author{Y.P. Jing} 
\affil{Shanghai Astronomical Observatory, the Partner Group of MPI f\"ur
Astrophysik, Nandan Road 80,  Shanghai 200030, China}
\author {Gerhard B\"orner} 
\affil {Max-Planck-Institut f\"ur Astrophysik,
Karl-Schwarzschild-Strasse 1, 85748 Garching, Germany}
\author{Yasushi Suto} 
\affil{Department of Physics and Research Center for the Early Universe, School of Science, University of Tokyo, Tokyo 113-0033, Japan.}
\begin{abstract}
  We report on the measurement of the two-point correlation function
  and the pairwise peculiar velocity of galaxies in the IRAS PSCz
  survey. The real space two-point correlation function can be fitted
  to a power law $\xi(r) = (r_0/r)^{\gamma}$ with $\gamma=1.69$ and
  $r_0=3.70 \mpc$. The pairwise peculiar velocity dispersion
  $\sigma_{12}(r_p)$ is close to $400 \kms$ at $r_p=3\mpc$ and
  decreases to about $150 \kms$ at $r_p \approx 0.2 \mpc$.  These
  values are significantly lower than those obtained from the Las
  Campanas Redshift Survey, but agree very well with the results of
  blue galaxies reported by the SDSS team later on. We have
  constructed mock samples from N-body simulations with a
  cluster-weighted bias and from the theoretically constructed GIF
  catalog. We find that the two-point correlation function of the mock
  galaxies can be brought into agreemnt with the observed result, but
  the model does not reduce the velocity dispersions of galaxies to
  the level measured in the PSCz data. Thus we conclude that the
  peculiar velocity dispersions of the PSCz galaxies require a biasing
  model which substantially reduces the peculiar velocity dispersion
  on small scales relative to their spatial clustering. The results
  imply that either the cosmogony model needs to be revised or the
  velocity bias is important for the velocity dispersion of the IRAS
  galaxies.
\end{abstract}

\section {Introduction}

Large catalogs of galaxies with their redshift and angular positions are
the major astronomical data sets which provide quantitative information
on the distribution and formation of the universe.  To properly decipher
this information requires the use of statistical tools, and to assess
their importance heavily relies on a quantitative comparison with
theoretical models.

In a recent series of papers (Jing, Mo, \& B\"orner 1998, hereafter JMB98;
Jing \& B\"orner 1998; Jing \& B\"orner 2001), various
statistical quantities from the Las Campanas Redshift Survey (LCRS)
have been determined including the two-point correlation function
(2PCF), $\xi(r)$, the power spectrum, $P(k)$, the pairwise peculiar
velocity dispersion (PVD), $\sigma_{12}(r)$, and the three-point
correlation function (3PCF), $\zetarr$. In addition, a detailed
comparison between these observational results and the predictions of
current cold dark matter (CDM) models has been carried out
with the help of mock samples constructed from N-body simulations.
It has been found that both the real-space 2PCF and the PVD can be measured
reliably from the LCRS, and that the observed 2PCF for LCRS is
significantly flatter than the mass 2PCF in CDM models on scales $\ls 1
\mpc$. The observed PVD also turned out to be lower than that of the
dark matter particles in these models. JMB98 introduced a
cluster-underweight (CLW) biasing model to account for these
discrepancies; CLW essentially assumes that the number of galaxies per
unit dark matter mass in a massive halo of mass $M$ decreases as
$\propto M^{-\alpha}$.  If $\alpha=0.08$, the 2PCF and the PVD of the
LCRS are well reproduced in a spatially-flat CDM model with
$\Omega_0^{0.6}\sigma_8 \approx 0.4$, where $\Omega_0$ is the 
density parameter of
the model and  $\sigma_8$ is the current rms linear density
fluctuation within a sphere of radius $8\mpc$.  

In this contribution, we report on the measurements of the 2PCF and PVD
of galaxies in another large redshift survey, the PSCz catalog, which
has become publicly available recently (Saunders et al. 2000). Since
galaxies in the PSCz survey are selected from the IRAS point source
catalog, they are supposed to be preferentially dominated by late
types in contrast to the LCRS galaxies. Thus the difference of those
clustering statistics between LRCS and PSCz should be interpreted as
an indication for the morphology-dependent biasing of galaxies, and we
will focus on the implications of the observed quantities for galaxy
formation theories. For the details of this work, we refer the readers
to our recent paper (Jing, B\"orner, \& Suto 2002).

\section{Clustering of PSCz galaxies and a comparision with the CDM predictions}
\begin{figure}
\plottwo{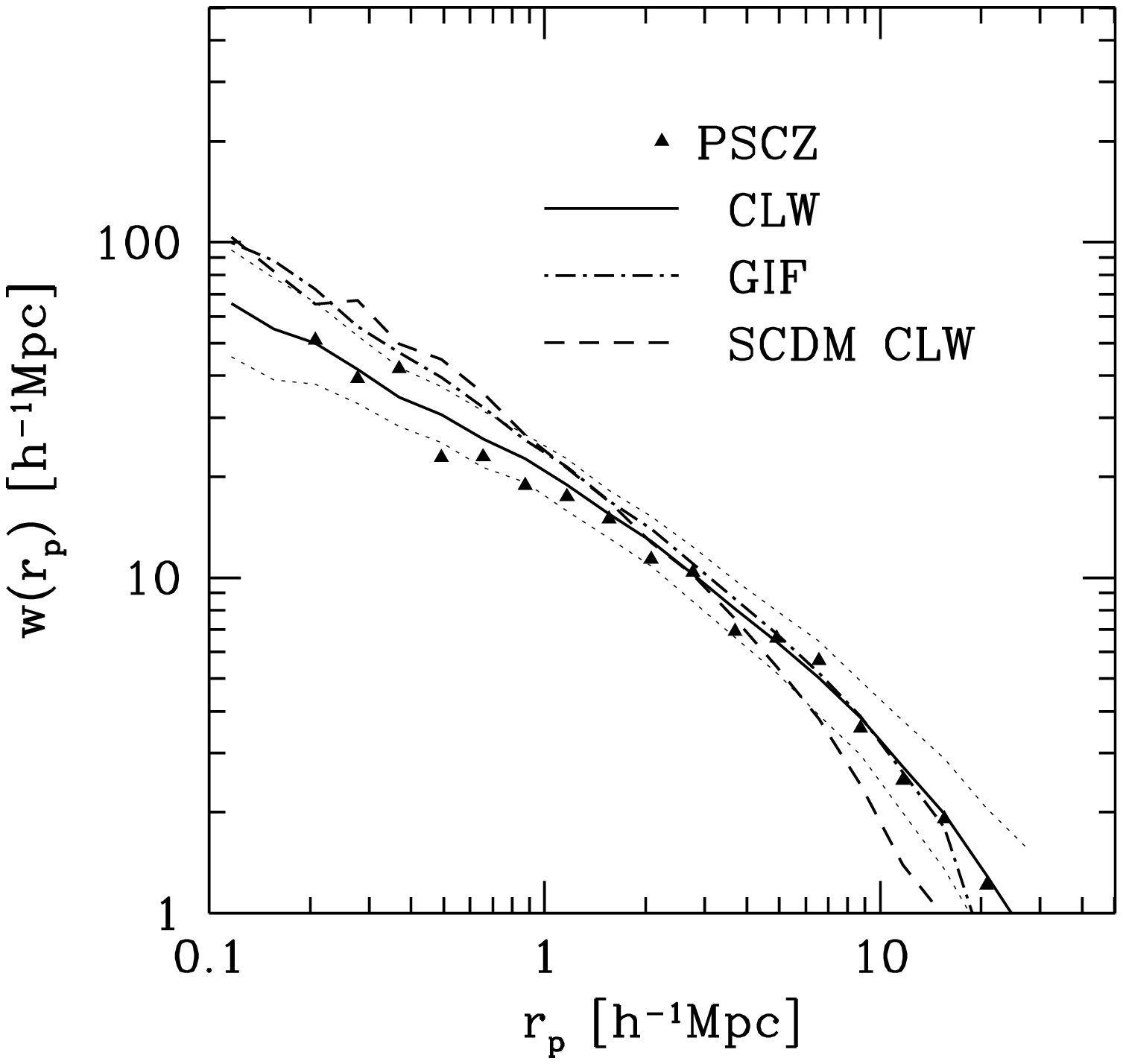}{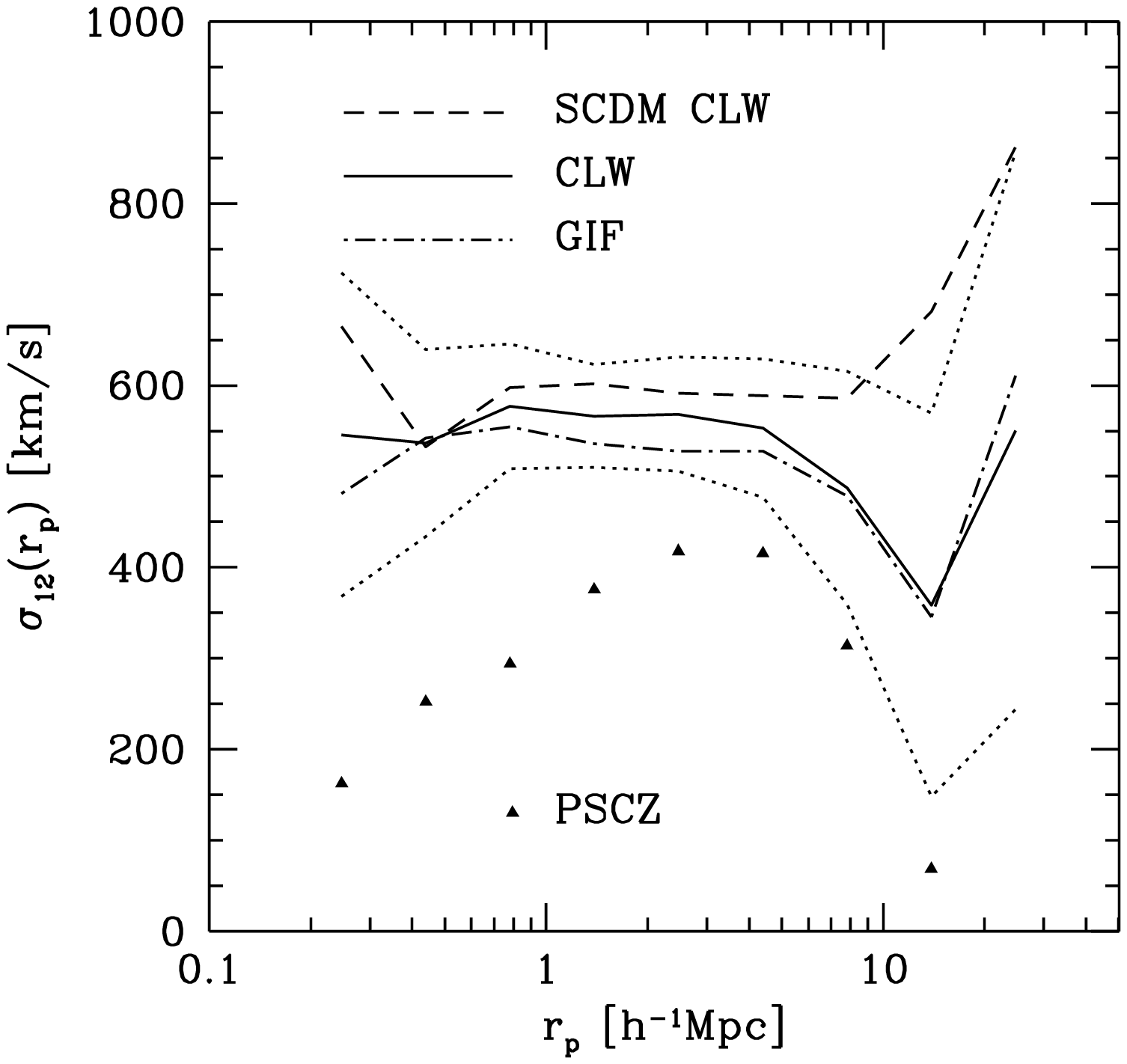}
\caption{ The predictions of CDM models vs the observation for the
projected two-point correlation function ({\it left}) and the pairwise
velocity dispersion ({\it right}). Triangles show the observational
results. The mean value and the $1\sigma$ limits predicted by the
cluster-weighted bias model are shown by the thick and thin lines
respectively, and the mean values of the GIF simulation and the SCDM
CLW model by the dot-dashed and dashed lines (without error bars). The
SCDM curve of the correlation function is shifted vertically by a
factor of $1/\sigma_8^2$ to account for the necessary linear bias in
this model.  }
\label{fig1}\end{figure}

We measure the two-point correlation function and the pairwise
velociity dispersion for the PSCz galaxies following exactly the
procedure of JMB98. The projected 2PCF $\wrp{}$ are presented in
Figure~1 with the filled triangles. The data can be well
fitted by a power law of $\xi(r) = (r_0/r)^\gamma$ with $r_0 = 3.70
\mpc$ and $\gamma =1.69$.  The slope is shallower and the amplitude is
lower than those for the LCRS (JMB98), reflecting the fact that the
PSCz galaxies are preferentially in the field environments.  The error
bars are not plotted for clarity but are very close to the error bars
predicted by the mock samples of the LCDM model with $\Omega_0=0.3$,
$\lambda_0=0.7$ and $\sigma_8=1.0$ (the dotted lines).  For the mock
samples, we have applied the CLW bias with $\alpha=-0.25$, which is
found to be in good agreement with the PSCz data (the solid line);
even the wiggly structure of the PSCz $\wrp{}$ below $0.8 \mpc$ is
recovered.

To follow up this point a bit more, we have also constructed mock
catalogs from the GIF simulation data (Kauffmann et al. 1999a,b).  From the
simulated catalogue, we select those galaxies with $\Delta V_{bg} \equiv
V_{b}-V_{g} \ge 1$, where $V_{b}$ and $V_{g}$ denote the V-band
magnitudes of the bulge and the whole galaxy.  This encompasses about 80
percent of the galaxies in the GIF catalog. The resulting $\wrp{}$ again
fits the observations quite well (Fig.~1) for $r_p \simgt
1\mpc$. While the amplitude of $\wrp{}$ for the GIF simulation is a bit
larger than the PSCz data for $r_p \simlt 1\mpc$, it still lies close to
the $+1 \sigma$ error line of our CLW mock samples.  So these
semi-analytic models of galaxy formation which incorporate physical
processes like star formation and supernova explosions in some global
way have a similar effect of reducing the number of galaxies per unit
dark matter mass as our simple bias prescription.

Figure~1 also displays the pairwise velocity dispersion for 
the self-similar infall model.
The PVD is much lower than that for the LCRS:
$\sigma_{12}$ just reaches $300 \kms$ at $r_p = 1 \mpc$, whereas
$\sigma_{12}(1 \mpc) = 570 \pm 80 \kms$ for the LCRS (JMB98). This is
again qualitatively consistent with the fact that spirals have smaller
random motions than the galaxies that reside in big clusters.
Comparing with the PVD for the CLW and GIF mock samples, we find that
the model predictions in all cases
significantly exceed the estimate for the PSCz.  Only around $r_p
\approx 3 \mpc$ the disagreement is not serious, especially if the
large error bars are taken into account.  In fact, we may speculate
that a CDM model with $\Omega_0 = 0.2$ may even produce quite a good
fit to the data at larger $r_p$, since the amplitude of the PVD scales
with $\sigma_8 \Omega_0^{0.6}$. Reducing the values of $\sigma_{12}$
accordingly, brings agreement on scales larger than $3 \mpc$. There
is, however, no way to reproduce the steep decrease towards small
values ($\sim 150 \kms$) at $r_p = 0.2 \mpc$ of the PSCz data. The
pairwise velocity dispersions are rather similar in the SCDM
($\Omega_0=1$ and $\sigma_8=0.6$) and in the LCDM models.

\section{Discussion and conclusions}
We have analyzed the data set of the IRAS PSCz galaxies
(Saunders et al. 2000), and computed the two-point correlation functions and
the pairwise peculiar velocity dispersion for these galaxies.  A
power-law fit to the real-space 2PCF $\xi(r) =(r_0/r)^\gamma$ gives an
exponent $\gamma=1.69$, and a amplitude $r_0=3.70 \mpc$. We show that
these results can be very well reproduced from mock samples
constructed for the LCDM model with the CLW bias of $\alpha =
0.25$. The bias needed for the PSCz galaxies is stronger than the LCRS
galaxies, since IRAS galaxies tend to avoid high-density cluster
regions. Mock smaples from the GIF simulation with an appropriate
choice of galaxies give similar 2PCF.

The pairwise peculiar velocity dispersion measured from the PSCz has a
much lower value, about $300 \kms$ at $r_p = 1 \mpc$, than the LCRS
result at that separation of $570 \pm 80 \kms$.  All the simulation
models which are consistent with the 2PCF of the PSCz predict
significantly larger values for PVD, although the CLW bias reduces the
PVD to within the $1 \sigma$ limit of this value, at least near $3
\mpc$ (if $\Omega_0=0.2$).  As discussed in the last section, the
decrease of $\sigma_{12}(r_p)$ for $r_p \ls 1\mpc$ for the PSCz data
is significant, and cannot be reproduced by the current simple models.
After our paper was submitted to the journal, we noted that the steep
decline of the PVD at the small speraration was also found in a recent
work of the SDSS team on clustering of blue galaxies (Zehavi et
al. 2002).  Thus, our results indicate that either the cosmogony model
needs to be revised or the velocity bias is important for the velocity
dispersion of the IRAS galaxies. We are investigating this issue with
high-resolution SPH/N-body simulations.

\acknowledgments 

The work is supported in part by the One-Hundred-Talent Program,
by NKBRSF(G19990754) and by NSFC(No.10043004) to YPJ.

\end{document}